\documentclass[aps, prd, twocolumn, showpacs, superscriptaddress, groupedaddress]{revtex4-1} 
\usepackage{graphicx}	
\usepackage{amssymb}
\usepackage{dcolumn}
\usepackage{color}
\usepackage{subfigure, rotating, bm, array}
\usepackage[pagebackref=false, colorlinks=true]{hyperref}
\hypersetup{linkcolor=blue, citecolor=blue,
urlcolor=blue} 

\begin{document}

\title{Tidal force effects and periodic orbits in null naked singularity spacetime}
\author{Siddharth Madan}
\email{siddharthmadan40@gmail.com}
\affiliation{International Center for Cosmology, Charusat University, Anand, GUJ 388421, India}
\author{Parth Bambhaniya}
\email{grcollapse@gmail.com}
\affiliation{International Center for Cosmology, Charusat University, Anand, GUJ 388421, India}
\date{\today}

\begin{abstract}
Naked singularities form during the gravitational collapse of inhomogeneous matter clouds. The final nature of the singularity depends on the initial conditions of the matter properties and types of matter profiles. These naked singularities can also be divided into two types: null-like and timelike singularities. The spacelike singularity of the Schwarzschild black hole can be distinguished from the null and timelike naked singularity spacetimes. In light of this, we investigate the precession of timelike bound orbits in the null naked singularity spacetime, as well as tidal force effects and geodesic deviation features. As a result, we find that the orbital precession of the timelike bound orbits in null naked singularity spacetime could be distinguished from the Schwarzschild precession case. The radial component of the tidal force has an intriguing profile, whereas the angular component has a profile which is comparable to that of a Schwarzschild black hole scenario. The geodesic deviation equation is then solved numerically, yielding results that resemble a Schwarzschild black hole. These characteristic features can then be used to discern amongst these singularities.

\end{abstract}
\maketitle

\section{Introduction}

The general relativistic effects of Einstein's gravity have been observed by various observations such as the great discoveries of the gravitational waves \cite{LIGOScientific:2016aoc}, the first-ever black hole shadow image of the M87 galactic center \cite{EventHorizonTelescope:2019dse}, stellar motions of the S-stars around the Milky-way galactic center \cite{GRAVITY:2020gka}, etc. These discoveries throughout have sparked a lot of interest in looking into the causal structure and the dynamics of spacetime around the galactic center.

The supermassive black hole and hence a spacetime singularity are thought to exist in the core regions of most galaxies, created by the catastrophic continuous gravitational collapse of primordial matter clouds. However, there is a fundamental mystery surrounding the singularity's causal structure. It is a well-established fact that there are three kinds of strong spacetime singularities in general relativity: spacelike, timelike, and null-like \cite{Dey:2020bgo,Joshi:2020tlq}. The spacelike singularity is causally detached from the other points of the spacetime manifold, and they may be hidden within horizons, whereas the null-like and timelike singularities are causally linked. 

According to a significant amount of research \cite{Dey:2020bgo,Joshi:2020tlq,Joshi:2011zm,Joshi:2013dva}, null-like and timelike singularities can be formed during the continual gravitational collapse of physically plausible matter clouds, and they contradict Roger Penrose's Cosmic Censorship Conjecture (CCC) \cite{Penrose:1969pc}. In \cite{Joshi:2020tlq}, authors propose a new spherically symmetric and static null naked singularity spacetime which is a solution of the Einstein field equations. It lacks a photon sphere, and they demonstrate that the singularity casts a shadow in the absence of a photon sphere. It has been determined that the event horizon and photon sphere are not essential for the formation of a shadow. From these studies, the existence of the upper bound of the effective potential of null-geodesics causes the formation of a shadow. Therefore, a spacetime can cast a shadow if its effective potential of null-geodesics has an upper bound. It has been shown that this null naked singularity spacetime satisfies strong, weak, and null energy conditions and verified that the Kretschmann scalar and Ricci scalar blow up at the center $r = 0$. Moreover, they point out that if any null geodesic emanates from the past null infinity, the photon would be infinitely red-shifted with respect to the asymptotic observer. 

However, if such type of null naked singularity exists in reality, then they must have distinct physical signatures. Many research studies have been published in which this topic has been investigated \cite{Joshi:2020tlq,
Joshi:2011zm}. One might come across abundant research literature concerning the aim to distinguish a naked singularity spacetime from any other spacetime which is engulfed by a null hypersurface or an event horizon. To distinguish these naked singularities, significant work has been done in the recent literature \cite{Bambhaniya:2019pbr,Joshi:2019rdo,Dey:2019fpv,Bambhaniya:2020zno,Bambhaniya:2021ybs,Bambhaniya:2021jum,Solanki:2021mkt,Dey:2020haf,Bambhaniya:2021ugr}, where authors investigate the shadows and orbital precession properties. Similarly, in this paper, we obtain an orbit equation of a test particle for the null-like naked singularity spacetime and we show the periastron precession of the timelike bound orbits for the same.   

Now, it is a well-known fact that a body in free-fall toward the center of another body is known to be stretched in the radial direction while being compressed in the angular direction. The stretching and compression are caused by gravity's tidal effect, which is caused by a variation in gravity's strength between two adjacent points \cite{MTW:1973,DInverno:1992,Carroll:2004,Hobson:2006}. The presence of tidal force phenomena is extremely common in our universe and has been a topic of popular scientific exploration for a major part of the twentieth century. To substantiate it further, John Wheeler proposed the possible disintegration of a star in the ergosphere of the Kerr black hole due to tidal interaction which results in the emission of jets formed from the debris of that star \cite{Goswami:2019fyk,Wheeler:1971}. Since then, the tidal disruption of stars in the presence of a massively gravitating object has been a topic for extensive research in the astrophysical research community \cite{Hills:1975,Carter:1982,Rees:1988,komossa:2015,Auchettl:2017,Rossi:2020rvv,Crispino:2016pnv,Shahzad:2017vwi,Hong:2020bdb}. 

In \cite{Crispino:2016pnv}, authors studied the effects of tidal force on a freely falling particle in Reissner-Nordstr\"om (RN) black hole and concluded that the radial and angular components of the tidal force change their signs at a boundary called the null hypersurface or the event horizon, which is not the case in the Schwarzschild black hole. In \cite{Gad2010}, the authors compared the tidal force effects in a stringy charged black hole with the Schwarzschild and the RN black holes. Shahzad and Jawad \cite{Shahzad:2017vwi} studied both the radial and angular components of the tidal force in the Kiselev black hole. In that paper, they considered the pervasive vicinity of radiation and dust fluids, also pointed out the change in sign between the event and the Cauchy horizons just as in the case of RN black holes. In \cite{Goel:2015zva}, authors examined the properties of tidal force in naked singularity spacetimes. The effects of tidal force have been analyzed for myriad other black holes \cite{Mahajan:1981,AbdelMegied:2004ni,Chan:1995fc,Nandi:2000gt,Cardoso:2012zn,Harko:2012ve,Uniyal:2014oaa,Sharif:2018a,Sharif:2018gzj,Junior:2020yxg,Junior:2020par}. Motivated by this, in this paper, we study the tidal force effects, geodesic deviation, and periodic orbits in the null naked singularity spacetime. We emphasize our results and compare them with the Schwarzschild black hole case.

The overview of this paper is given as: In the Section (\ref{II}), we have discussed the intricacies of this spacetime i.e. a brief on the preceding work on this metric and its characteristics. In Section (\ref{III}), we check the periodic orbits for a particle by varying a couple of parameters. In the section (\ref{IV}), we present the calculations for tidal force effects in this spacetime. We have been able to obtain some riveting results with the aid of tetrad formalism for the radial component of tidal force in this metric. In section (\ref{V}), we have compared the numerical solution of the geodesic deviation equation for the null naked singularity metric with the Schwarzschild black hole metric. Conclusions are drawn in Section (\ref{VI}). Throughout this paper, we consider gravitational constant (G) and speed of light (c) as unity. 
\section{Null Naked Singularity Spacetime}
\label{II}
\begin{figure}
\centering
\subfigure[]
{\includegraphics[scale=0.55]{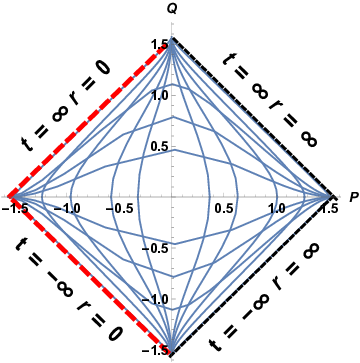}}
\hspace{0.2cm}

 \caption{Penrose diagram of null naked singularity}
 \label{nullpenrose}
\end{figure}
The line element representation for any geometry can be written as,
\begin{equation}
d{s}^2 =g_{\mu\nu}dx^{\mu}dx^{\nu},
\end{equation}
this can be elaborated further by delineating all the summed over components as,
\begin{equation}
    d{s}^2=-f(r)dt^2 + f(r)^{-1}dr^2 +
r^2(d\theta^2 + \sin^2\theta d\phi^2),
\label{eq2}
\end{equation}
and the metric for null naked singularity case can be given by \cite{Joshi:2020tlq},
\begin{equation}
ds^2 = -\frac{dt^2}{\left(1+\frac{M}{r}\right)^2}+\left(1+\frac{M}{r}\right)^2dr^2 +  +r^2d\Omega^2\,\, , 
\label{eq2.3}
\end{equation}
where, $d\Omega^2=d\theta^2+\sin^2\theta d\phi^2$ and $M$ is the Arnowitt-Deser-Misner (ADM) mass of the above spacetime. Now from the above two equations (\ref{eq2}) and (\ref{eq2.3}) we can write,

\begin{equation}
f(r) =  \left(1+\frac{M}{r}\right)^{-2}, 
\end{equation}
now one can give the binomial expansion for $f(r)$ as,
 \begin{equation}
f(r) =  \left[1-\frac{2M}{r}+3\left(\frac{M}{r}\right)^{2}-...\right], 
\end{equation}
from this binomial expansion, it is evident that in the large r limit, this metric mimics the Schwarzschild metric and in asymptotically infinite limit the metric behaves akin to the Minkowski spacetime. Even though the metric resembles the Schwarzschild metric at a large distance, near the singularity, the causal structure of this spacetime becomes different from the causal structure of Schwarzschild spacetime. 
It is shown in \cite{Joshi:2020tlq} that from the expressions of Kretschmann scalar and Ricci scalar at the center $(r = 0)$ there exists a strong curvature singularity and no null surfaces such as the event horizon engulfs the singularity, hence the name naked singularity. Fig. \ref{nullpenrose} shows the Penrose diagram of the null naked singularity spacetime. The most intriguing information to look for in this diagram is that any null geodesic emanating from the past null infinity would allow the photon to be infinitely red-shifted in regards to the asymptotic observer \cite{Bambhaniya:2021jum}. As a result, at r = 0, it is a nulllike naked singularity.
There it is also shown that this metric satisfies all energy conditions. Using Einstein field equations, the expressions for the energy density and pressures of this naked singularity spacetime are obtained. Eventually, it is deduced that this spacetime is seeded with an anisotropic fluid for which the equation of state parameter tends to -1/3 when $r\rightarrow{0}$ and it approaches 1/3 as $r\rightarrow{\infty}$.
\begin{figure*}
\centering
\subfigure[Timelike orbits in null-like naked singularity spacetime for $h=4$, where $r_{min}=8.17$ and $E=-0.008$]
{\includegraphics[scale=0.57]{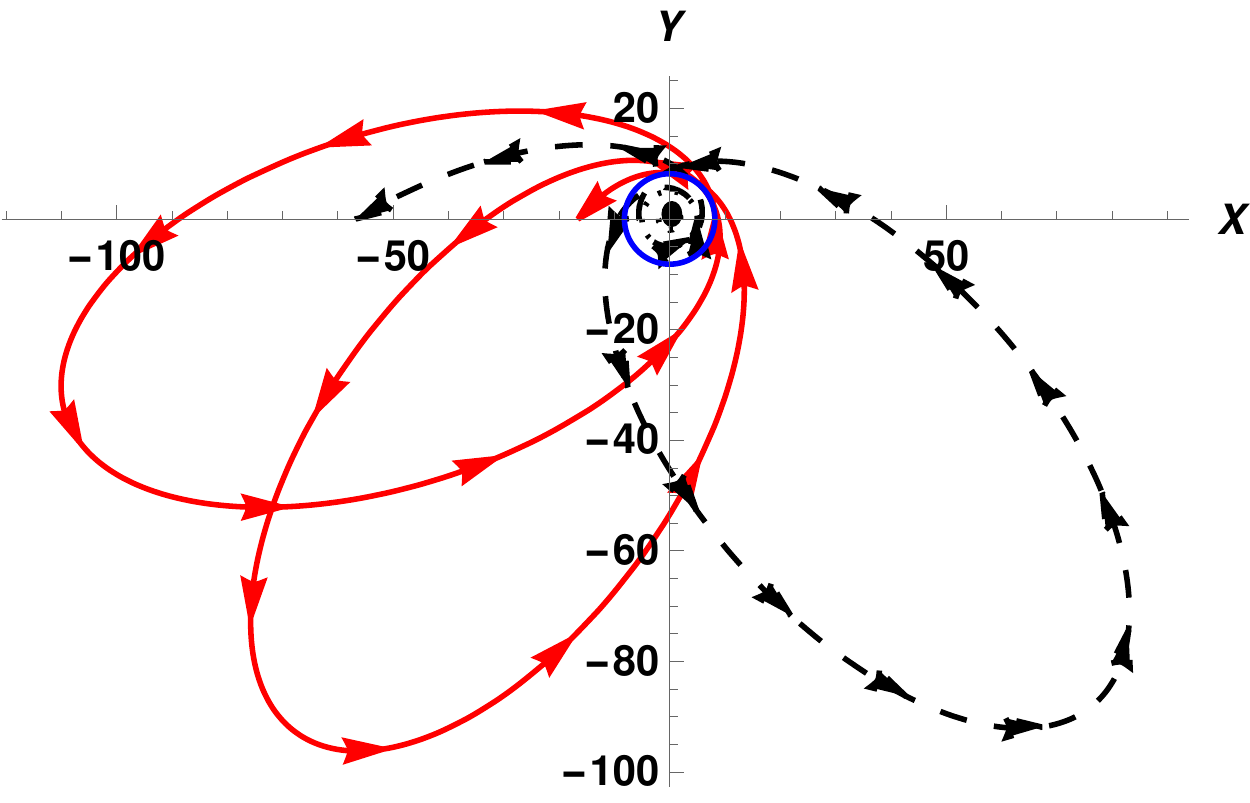}}
\hspace{0.2cm}
\subfigure[Timelike orbits in null-like naked singularity spacetime for $h=5$, where $r_{min}=13.74$ and $E=-0.008$]
{\includegraphics[scale=0.57]{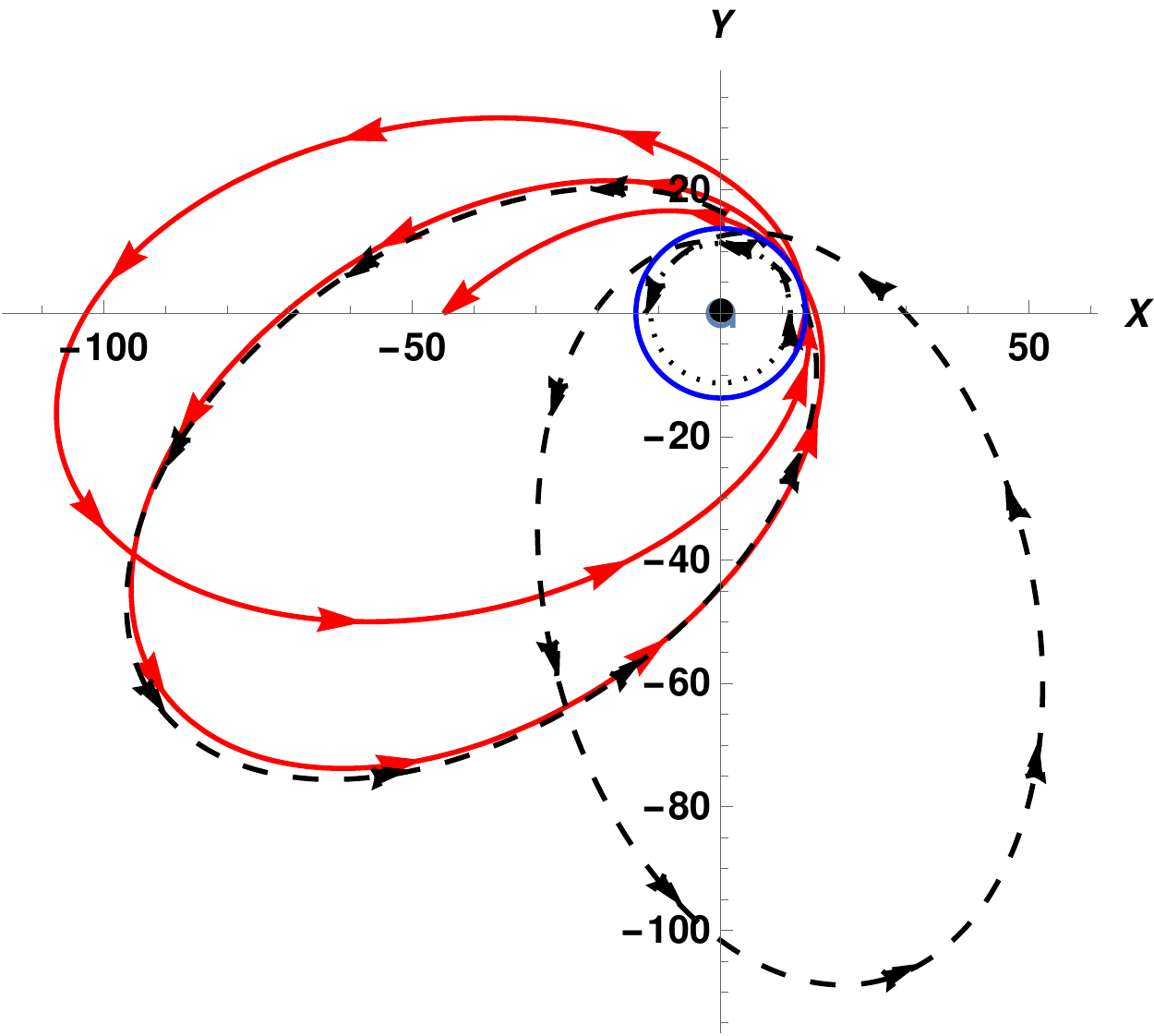}}
\hspace{0.2cm}
\subfigure[Timelike orbits in null-like naked singularity spacetime for $h=6$, where $r_{min}=21.58$ and $E=-0.008$]
{\includegraphics[scale=0.55]{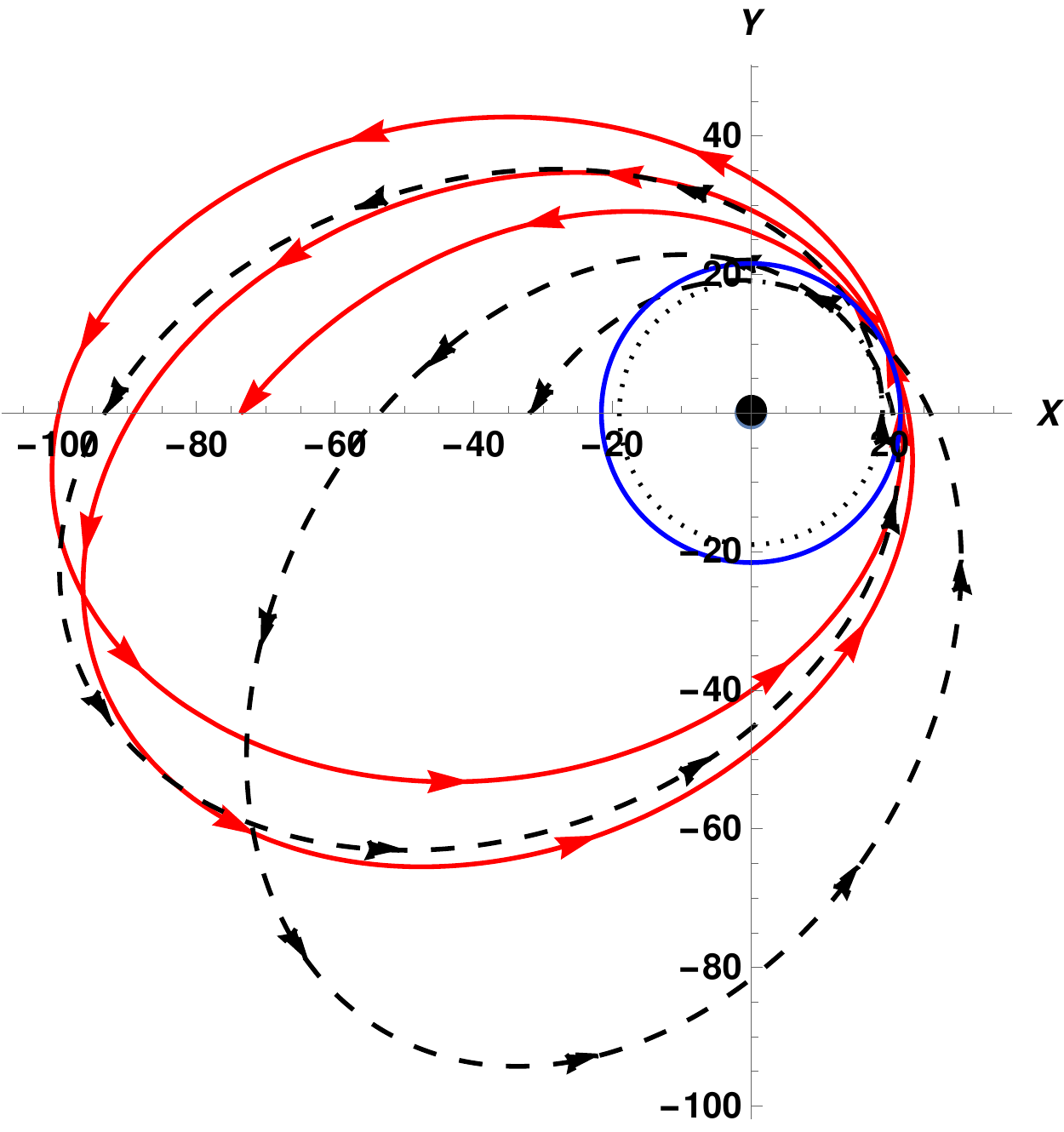}}
\hspace{0.2cm}   
\subfigure[Timelike orbits in null-like naked singularity spacetime for $h=7$, where $r_{min}=33.55$ and $E=-0.008$]
{\includegraphics[scale=0.55]{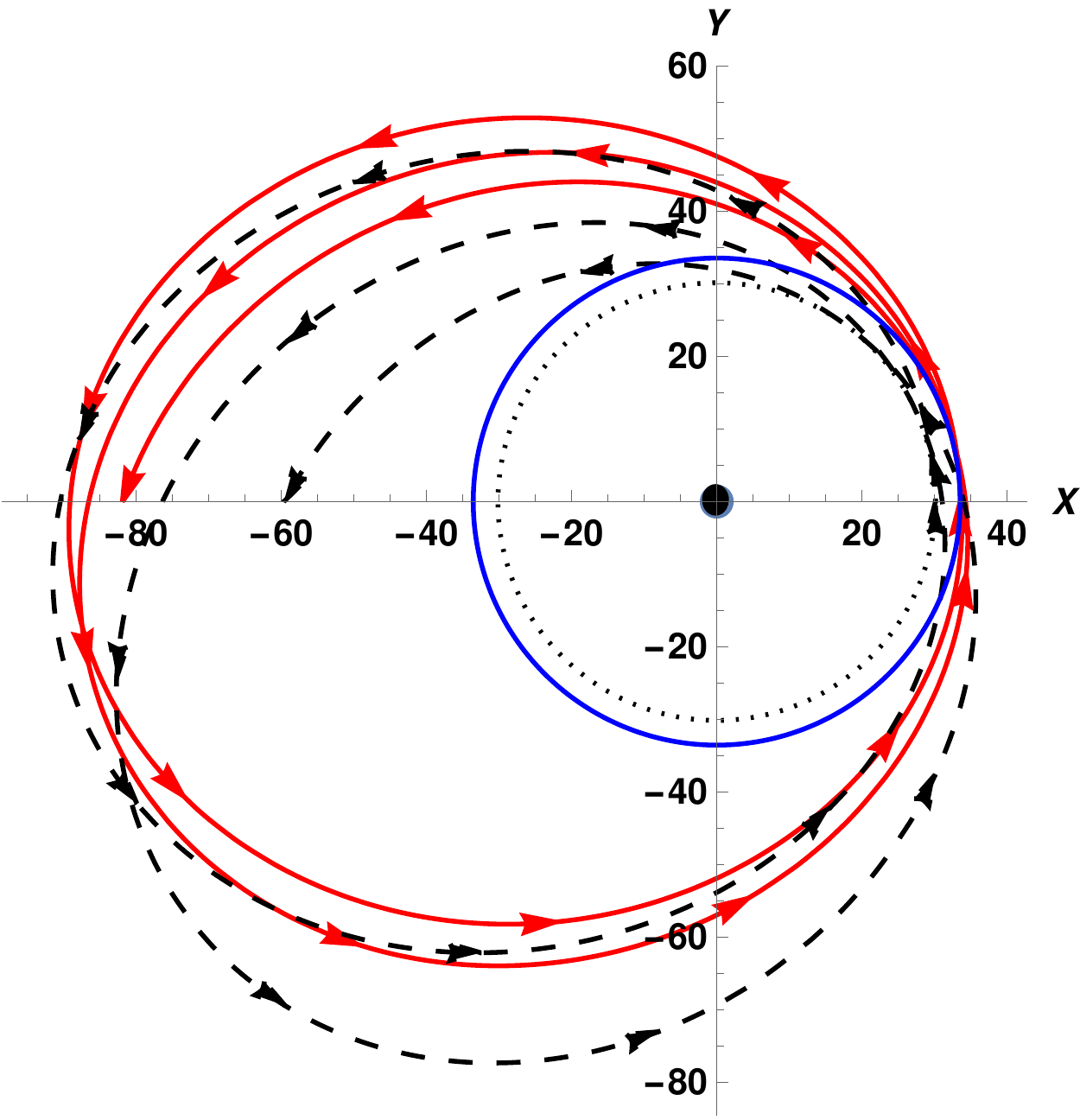}}
\hspace{0.2cm}

 \caption{In this figure, relativistic orbits (The red lines) of a test particle in the null-like naked singularity spacetime are shown. The blue circle represents the minimum approach of a test particle (or periastron points) in a naked singularity case. While dotted black orbits indicate the orbital precession in the Schwarzschild black hole spacetime. The dark black region at the center defines the Schwarzschild radius. where the total mass $M=1$.}
 \label{precgen}
\end{figure*}

\section{Periodic orbits}
\label{III}
No material particle or light may escape to distant observers if the singularity is surrounded by a null hypersurface (event horizon or trapped surface). This singularity is called a space-like singularity, since it is not causally connected to any other spacetime points. Other sorts of singularities can also develop as a result of the gravitational collapse of an inhomogeneous matter cloud \cite{joshi,mosani3,mosani4,Joshi:2011zm,JNW}. These gravitationally strong singularities are linked to other spacetime points, which are referred to as null and time-like singularities \cite{Dey:2020bgo,Joshi:2020tlq}. Now, the conserved energy ($\gamma$) and angular momentum ($h$) per unit rest mass confirm the axial and temporal symmetries in the null singularity spacetime given in equation (\ref{eq2.3}). Therefore, we can write the expressions of these conserved quantities as, 
\begin{equation}
    \gamma = -g_{tt}(r) u^t =\frac{1}{\left(1+\frac{M}{r}\right)^2}\left(\frac{dt}{d\tau}\right)\,\, ,\,\,\,
   \label{congen}
\end{equation}
\begin{equation}
    h = g_{\phi\phi}(r) u^{\phi}=r^2\left(\frac{d\phi}{d\tau}\right)\,\,,
\end{equation}
where, $\tau$ is the proper time of the particle. Using these symmetries property and normalisation of the four velocity conditions for the timelike geodesics ($u^\alpha u_\alpha=-1$), we can determine the total relativistic energy for $\theta=\pi/2$ as, 
\begin{equation}
    E=\frac{1}{2}\left[\left(\frac{dr}{d\tau}\right)^2+V_{eff}(r)\right],
\end{equation}
\noindent
where, $V_{eff}$ is the effective potential of the null naked singularity spacetime, which is derived as,
 \begin{equation}
    V_{eff}(r)=\frac{\left(\frac{h^2}{r^2}+1\right)}{\left(1+\frac{M}{r}\right)^2}-1.
\end{equation}
Note that, we have considered the timelike bound orbits only in the equatorial plane $(\theta=\pi/2)$, since a spacetime metric is spherically symmetric and static. Moreover, it is a widely acknowledged fact that the inner most stable circular orbit (ISCO) could be obtain by solving $\frac{dV_{eff}}{dr}=0$ for the extremum value of $r$. Therefore, one can find ISCO for Schwarzschild black hole case at $r_{isco}=6M$. While, in the null naked singularity case, we obtain expression of ISCO to be at $r={h^2}/{M}$. This expression explicitly demonstrates that ISCO for the case of null naked singularity spacetime extends up to the singularity at $r=0$ as $h\rightarrow{0}$. This means that for any particle with some angular momentum, the radius of its inner most stable circular orbit would depend directly on the square of its own angular momentum and inversely proportional to the mass of that singularity. It is also evident that for a particle with zero angular momentum there would be no ISCO and it would eventually plunge into the singularity. Moving on, the timelike bound orbits can be accomplished by satisfying the conditions, 
\begin{equation}
   E - V_{eff}(r)>0; \, \, \  V'_{eff}(r)=0; \, \, \  V''_{eff}(r)>0,
\end{equation}
where, prime denote for the derivative with respect to `r'. Using these timelike bound orbit conditions, we can derived an expression for the shape of an orbit as,
\begin{equation}
    \left(\frac{dr}{d\phi}\right)^2=\frac{r^4}{h^2}\left[\gamma^2-\frac{\left(\frac{h^2}{r^2}+1\right)}{\left(1+\frac{M}{r}\right)^2}\right],
    \label{shape}
\end{equation}
this above expression implies how the coordinate `r' change as the coordinate `$\phi$' change. Now, one can derive the orbit equation for null naked singularity metric by differentiating the above Eq. (\ref{shape}) with respect to $\phi$ and we get,
\begin{equation}
    \frac{d^2u}{d\phi^2}-\frac{M(1+h^2u^2)}{h^2(1+Mu)^3}+\frac{u}{(1+Mu)^2}=0,
\end{equation}
the equation above is known as the orbit equation of a test particle for null naked singularity spacetime. It is considerably difficult to solve this equation analytically since it is second order non-linear differential equation. Therefore, we solve it numerically and shown in the Fig. \ref{precgen}. 

In Fig. \ref{precgen}, we show that the periastron precession of the timelike bound orbits for different angular momentum values (h = 4, 5, 6, 7) in the null naked singularity spacetime (The red orbits). The blue circle represents the minimum approach of a test particle (or periastron points) in a naked singularity case. While dotted black orbits indicate the orbital precession in the Schwarzschild black hole spacetime. The dark black region at the center defines the Schwarzschild radius. One can see from Fig. \ref{precgen}, that the precession angle would decrease as the angular momentum of a test particle increases, and hence the distance of the minimum approach will increase. In Schwarzschild spacetime, the precession angle of the orbit is larger than the precession angle of the orbit in naked singularity spacetime. However, at large distances from the singularity, the precession angle of the orbit would be nearly similar as we increase the angular momentum of the particle `h'.

\section{Tidal force effects}
\label{IV}
Now in this section, we dive into the intricacies of the equations involving tidal forces in our spacetime i.e. null naked singularity. In order to examine the equation for the distance between two infinitesimally close and free falling particles, we take the equation for the spacelike components of geodesic deviation vector $\eta^{\mu}$ to be:
\begin{equation}
    \frac{D^2 \eta^\mu}{D\tau^2} - R^{\mu}_{\nu\rho\sigma}v^{\nu}v^{\rho}\eta^{\sigma}=0,
\end{equation}
where $R^{\mu}_{\nu\rho\sigma}$ is the Riemann curvature tensor and $v^{\mu}$ is the unit tangent vector to the geodesic \cite{Hobson:2006}. In order to solve for the Riemann curvature tensor we take the aid of tetrad formalism. Tetrads are geometric objects that are orthonormal and form a set of local coordinate bases i.e. a locally defined set of four linearly independent vector fields called tetrads or vierbien \cite{De}. At every point on some geodesic, there is a tetrad frame that forms a local inertial reference frame which is valid in the region nearby. In this region, the laws of special relativity apply. An orthonormal basis, unless it is also a coordinate basis, does not have enough information to provide the line element (or the connection). There can be infinitely many orthonormal bases at a particular point related to each other by Lorentz transformations. But to determine those orthonormal bases which carry the information of the metric, we must find a linear transformation from the orthonormal basis to a coordinate basis:
\begin{equation} \label{tetrad01}
    \vec{e}_{\mu} = \hat{e}^{\hat{\mu}}_{{\mu}} \vec{e}_{\hat{\mu}}
\end{equation} 
where, $\vec{e}_{\mu}$ represents the coordinate basis, $\vec{e}_{\hat{\mu}}$ represents the orthonormal basis and $\hat{e}^{\hat{\mu}}_{{\mu}}$ are the coefficients also known as the tetrad components. These tetrad components are the coefficients that contain the information of a metric and eventually enable these orthonormal basis to extract physical, measurable quantities from geometric, coordinate-free objects in curved manifolds. These tetrads may be inverted in the obvious ways:
\begin{equation}
    \vec{e}_{\hat{\mu}} = \hat{e}_{\hat{\mu}}^{{\mu}} \vec{e}_{\mu} 
\end{equation}
where, $\hat{e}_{\hat{\mu}}^{{\mu}}\hat{e}_{\nu}^{{\hat{\mu}}}=\delta_{\nu}^{\mu}$. The metric components in the coordinate basis follow from the tetrad components:
\begin{equation} \label{tetrad02}
    g_{\mu\nu} = \vec{e}_{\mu} . \vec{e}_{\nu} = \eta_{\hat{\mu}\hat{\nu}} \hat{e}^{\hat{\mu}}_{{\mu}}\hat{e}^{\hat{\nu}}_{{\nu}}
\end{equation}
The equation above is the key result allowing us to use these orthonormal bases in curved spacetime. With the aid of this formalism the parallel transport of tangent vectors along the geodesics can be convened and hence, this tetrad formalism satisfies geodesic equations in that curved spacetime \cite{Bert:2002}. The tetrad components for freely falling reference frames can be obtained from Eq. (\ref{tetrad01}) and Eq. (\ref{tetrad02}) to be:
\begin{equation}
    \hat{e}^{\mu}_{\hat{0}}\equiv\left\{\frac{E}{f},-\sqrt{E^{2}+f},0,0\right\},
\end{equation}
  \begin{equation}
      \hat{e}^{\mu}_{\hat{1}}\equiv\left\{-\frac{\sqrt{E^{2}+f} }{f},E,0,0\right\},
  \end{equation}  
\begin{equation}
    \hat{e}^{\mu}_{\hat{2}}\equiv\left\{0,0,\frac{1}{r},0\right\},
\end{equation}
\begin{equation}
    \hat{e}^{\mu}_{\hat{3}}\equiv\left\{0,0,0,\frac{1}{r\sin{\theta}}\right\},
\end{equation}
where, $(x^0,x^1,x^2,x^3) = (t,r,\theta,\phi)$. These tetrad components satisfy the following orthonormality condition,
\begin{equation}
    \hat{e}^{\ \alpha}_{\hat{\mu}}\hat{e}_{\hat{\nu} \alpha} = \eta_{\hat{\mu}\hat{\nu}},
\end{equation}
where, $\eta_{\hat{\mu}\hat{\nu}}$ is the Minkowski metric \cite{DInverno:1992}.
We have that $\hat{e}^{\mu}_{\hat{0}}=v^\mu$.
The geodesic deviation vector, also called the separation vector, can be expanded as,
\begin{equation}
    \eta^{\mu} = \hat{e}_{\hat{\nu}}^{\ \mu}\eta^{\hat{\nu}}.
\end{equation}
Note that for a fixed temporal component $\eta^{\hat{0}}=0$   \cite{DInverno:1992}. Therefore, subsequently one can obtain non-vanishing independent components of the Riemann tensor in spherically symmetric spacetimes, including the null naked singularity spacetime as \cite{Wald},

\begin{figure*}
\centering
\subfigure[]
{\includegraphics[scale=0.45]{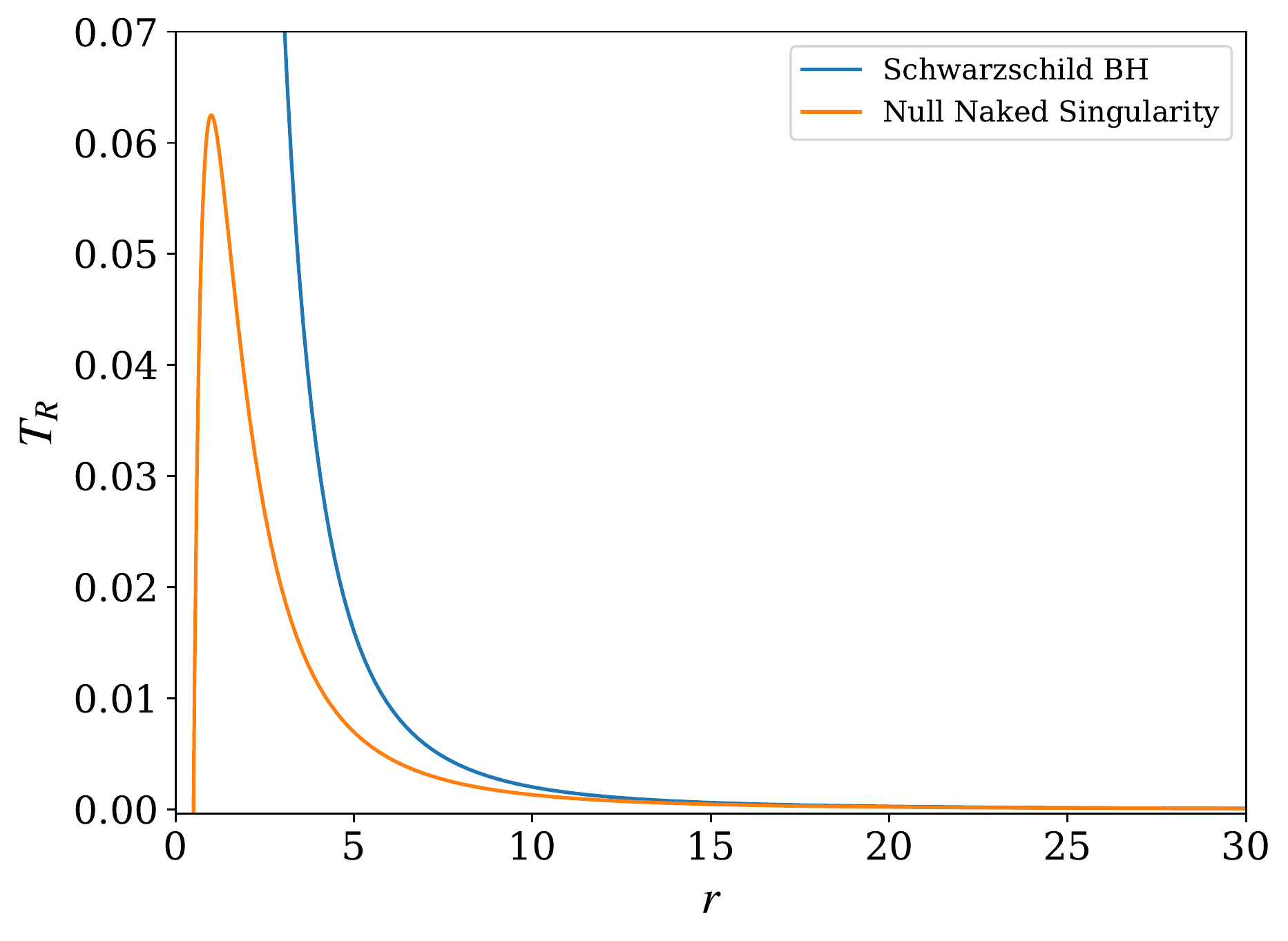}}
\hspace{0.2cm}
\subfigure[]
{\includegraphics[scale=0.45]{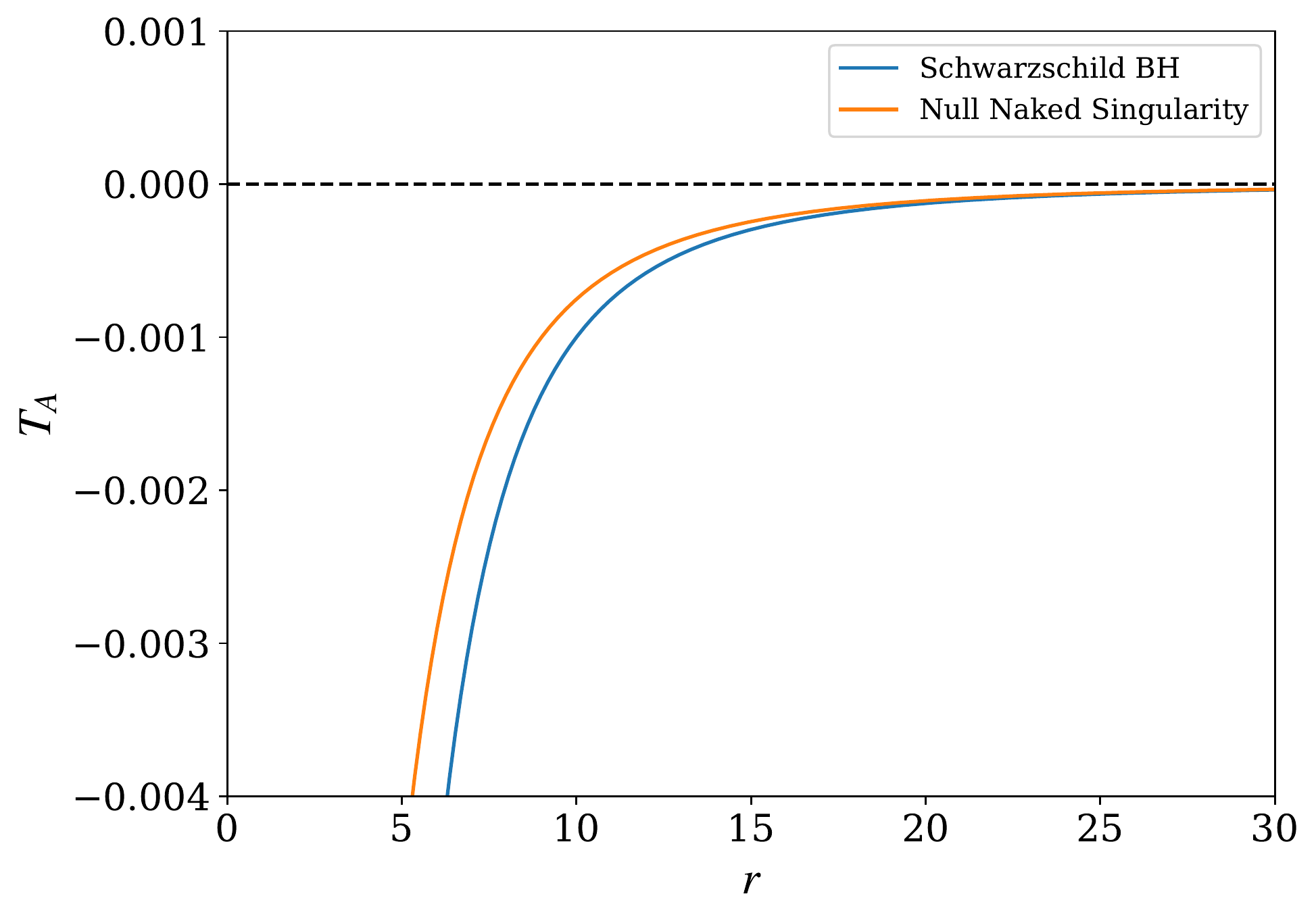}}
\hspace{0.2cm}
\caption
{(a) Radial tidal effects, $T_R = \frac{1}{\eta^{\widehat{1}}}\frac{d^{2}\eta^{\widehat{1}}}{d\tau^2}$ and (b) Angular tidal effects, $T_A=\frac{1}{\eta^{\widehat{i}}}\frac{d^{2}\eta^{\widehat{i}}}{d\tau^2}$ for radially freely falling particle in the null naked singularity and Schwarzschild black hole spacetimes.
}\label{tidal_plt}
\end{figure*}

\begin{figure}
    \centering
    \includegraphics[scale=0.28]{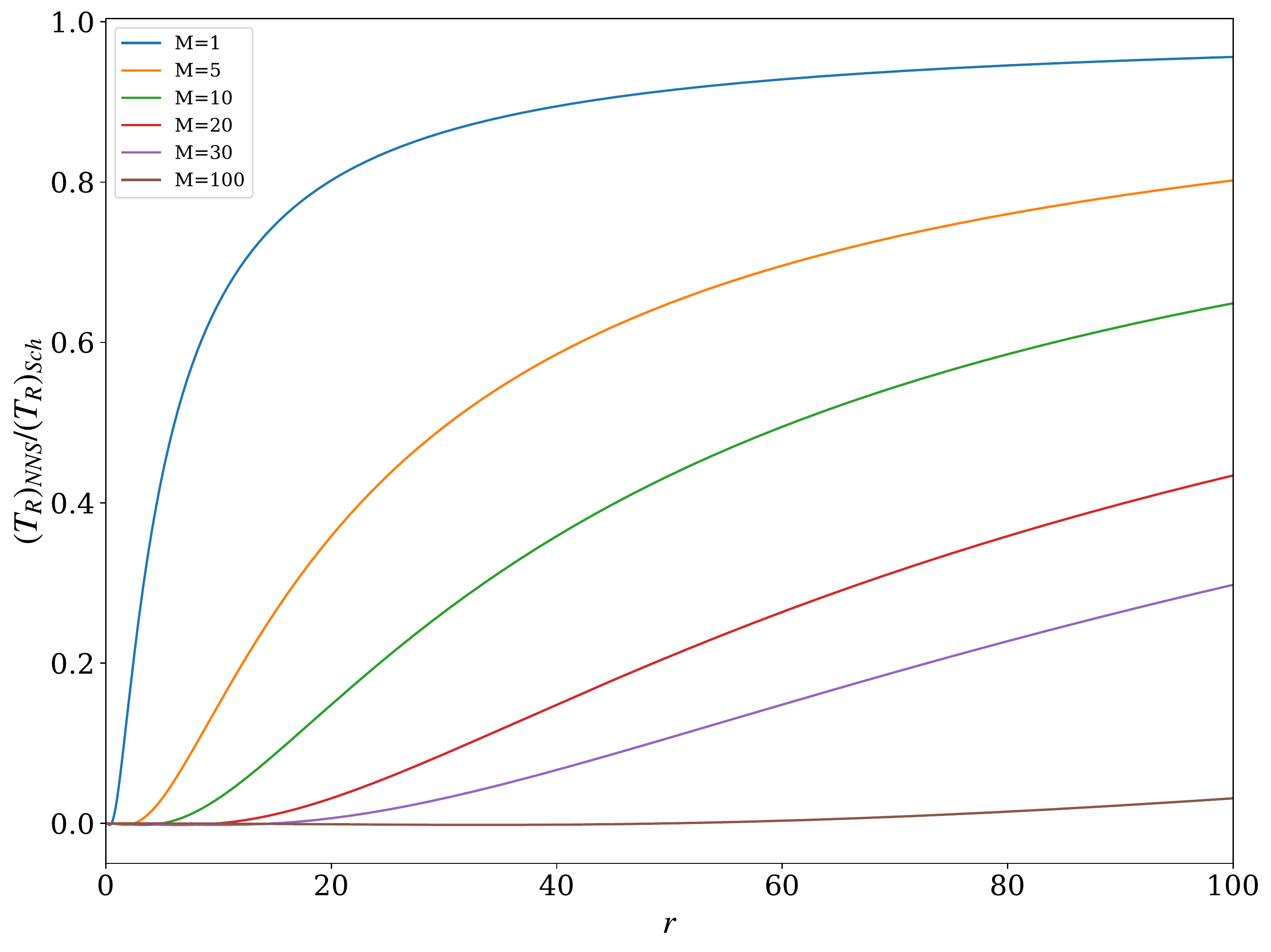}
\caption
{Radial tidal force ratio for the null naked singularity spacetime $(T_R)_{NNS}$ over Schwarzschild black hole singularity spacetime $(T_R)_{Sch}$.}
    \label{tidal_comp}
\end{figure}
\begin{equation}
 R^1_{\ 212}= -\frac{rf^{\prime}}{2}\,\, ; \ 
 R^1_{\ 313}= -\frac{rf^{\prime}}{2}\sin^2\theta,
\end{equation}
\begin{equation}
 R^1_{\ 010}= \frac{ff^{\prime \prime}}{2}\,\, ; \
 R^2_{\ 323}= \left(1-f\right)\sin^2\theta,
\end{equation}
\begin{equation}
 R^2_{\ 020}= \frac{ff^{\prime}}{2r}\,\, ; \  R^3_{\ 030}= \frac{ff^{\prime}}{2r}.
\end{equation}
Now using the tetrad formalism, we obtain the Riemann curvature tensor in terms of tetrad basis to be:
\begin{equation}\label{eq2.4}
R^{\hat{\alpha}}_{\hat{\beta}\hat{\gamma}\hat{\delta}}=R^{\mu}_{\nu\rho\sigma}\hat{e}_{\mu}^{\hat{\alpha}}\hat{e}_{\hat{\beta}}^{\nu}\hat{e}_{\hat{\gamma}}^{\rho}\hat{e}^{\sigma}_{\hat{\delta}}.   
\end{equation}
Upon obtaining the expressions from Eq.~(\ref{eq2.4}) while considering that the vectors ${\hat{e}_{\hat{\nu}}}^{\ \mu}$ are parallelly transported along the geodesic, one can find the following equations for tidal forces in radial free-fall reference frames \cite{Crispino:2016pnv},
\begin{equation}\label{radial}
   \ddot{\eta}^{\widehat{1}} = -\frac{f^{\prime \prime}}{2}\eta^{\widehat{1}}, 
\end{equation}
 \begin{equation}\label{angular}
   \ddot{\eta}^{\widehat{i}} = -\frac{f^{\prime}}{2r}\eta^{\widehat{i}},
 \end{equation}
where $i=2,\,3$. Now, we can plug in the second and the first derivatives of $f$ with respect to $r$ in Eq. (\ref{radial}) and Eq. (\ref{angular}) respectively to obtain,
\begin{equation}\label{radial02}
   \ddot{\eta}^{\widehat{1}}  =  -\frac{M(M-2r)}{(M+r)^4}\eta^{\widehat{1}}, 
\end{equation}
 \begin{equation}\label{angular02}
   \ddot{\eta}^{\widehat{i}}  =  -\frac{M}{(M+r)^3}\eta^{\widehat{i}}.
 \end{equation}
 The above two equations Eq. (\ref{radial02}) and Eq. (\ref{angular02}) are the tidal force equations for a radially free-falling frame in the null naked singularity metric, where Eq. (\ref{radial02}) represents the radial component of the tidal force and Eq. (\ref{angular02}) represents the angular component of the same. Using these two equations, one can generate two separate plots, one each for radial and angular components of the tidal force as shown in Fig. \ref{tidal_plt}. The process of compression in the angular direction and stretching in the radial direction is known as 'spaghettification'.
 
 \begin{figure*}
\centering
\subfigure[]
{\includegraphics[scale=0.45]{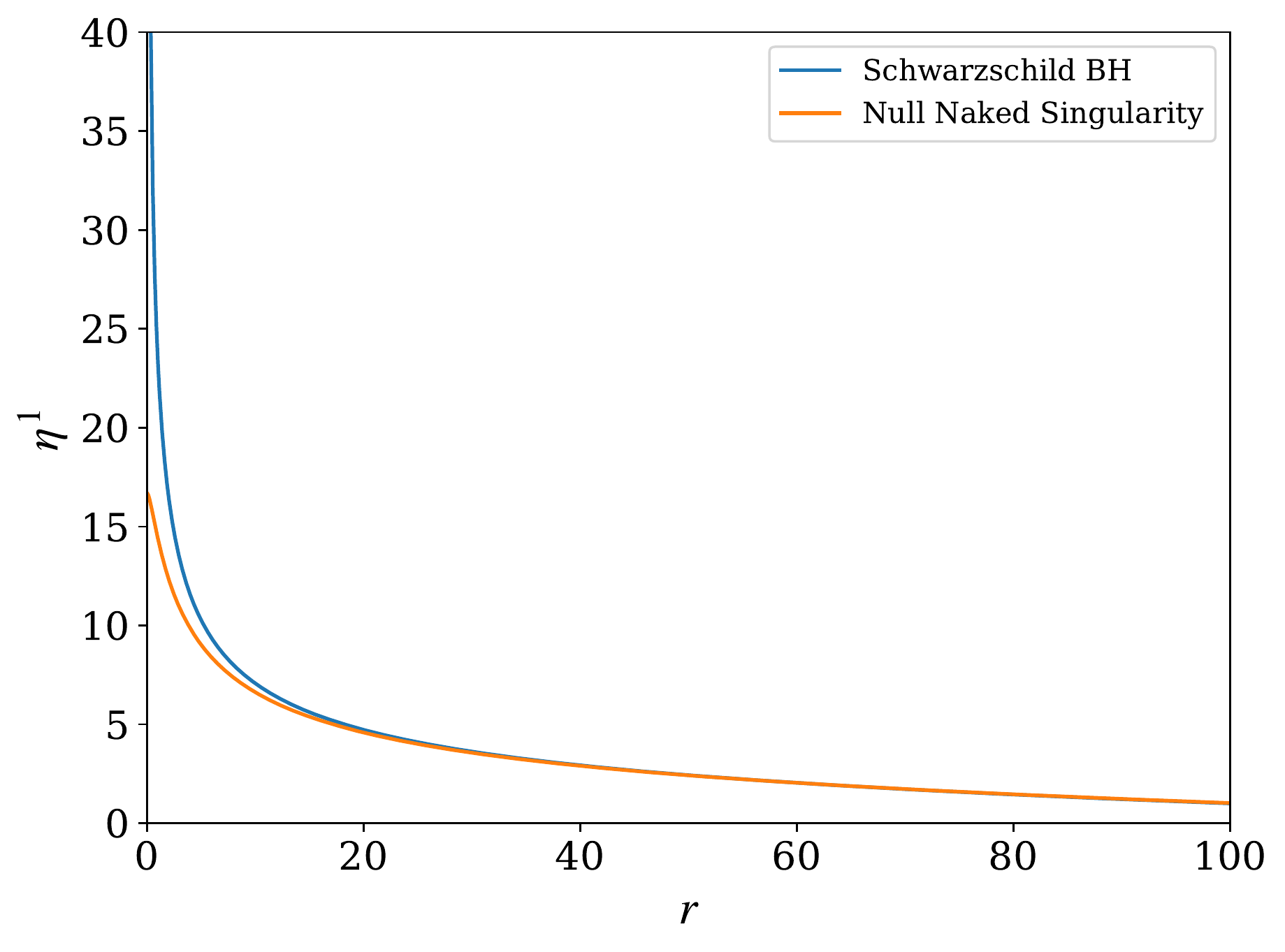}}
\hspace{0.2cm}
\subfigure[]
{\includegraphics[scale=0.45]{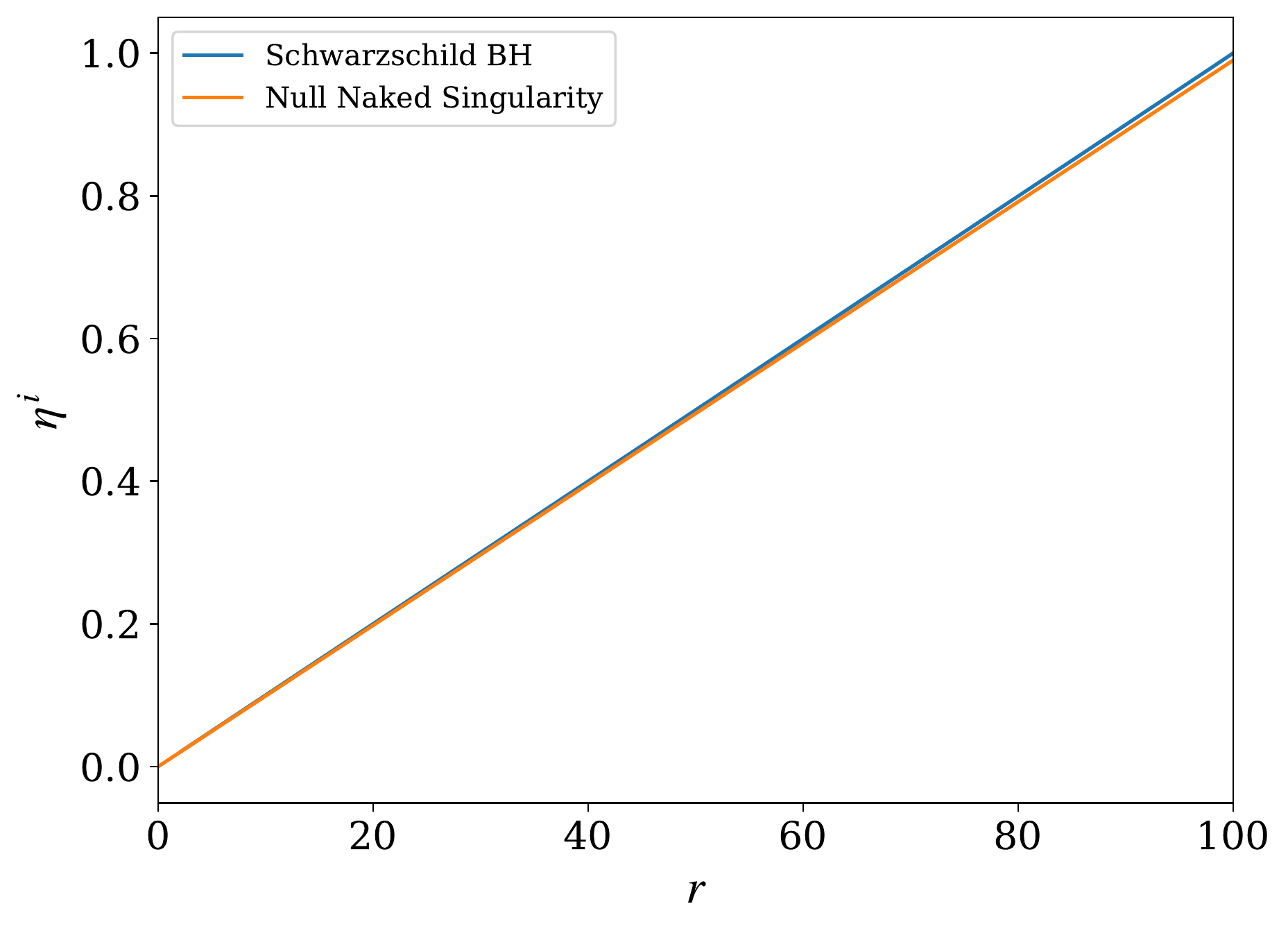}}
\hspace{0.2cm}
\caption
{(a) Radial solutions of the geodesic deviation equation for
$\eta^{\hat 1}(b) = 1$, $\frac{d\eta^{\hat 1}(b)}{d\tau} = 0$ with
$M = 1$, 
(b) Angular solutions of the geodesic deviation equation for $\eta^{\hat i}(b)=1$, $\frac{d\eta^{\hat
i}(b)}{d\tau}=0$, $M = 1$.}\label{geodesic}
\end{figure*}
 
 The radial component of the tidal force as described in Fig. \ref{tidal_plt}.(a) shows clear differences between the null naked singularity and the Schwarzschild black hole that can be summarised as follows. One can see from this figure that the radially free-falling object will stretch initially as it approaches the singularity at $r=0$. At a finite distance from the singularity, the signature of the radial tidal force flips and eventually behaves like a compressing force. This can also be verified from Eq. (\ref{radial02}) where the sign of the radial tidal force remains positive when $2r>M$ and at $r=M$ the radial tidal force reaches its maximum value. And beyond this maxima i.e. when $2r<M$ the radial tidal force component becomes negative and as $r\rightarrow{}0$ the radial tidal force converges to a finite value which is equivalent to $-1/M^2$. This indicates no more stretching of the radially infalling object beyond this finite maxima and rather the object gets subjected to finite compression until the plunge into the singularity. This result is contrary to the Schwarzschild black hole case where the radially infalling object is subjected to infinite stretching as it approaches the singularity. 
 
 For the angular component of the tidal force as shown in Fig. \ref{tidal_plt}.(b), the rate of change of angular tidal force for both cases is distinguishable from one another. Interestingly, in this case, the angular tidal force approaches $-1/M^2$ as $r \rightarrow{}0$ from Eq. (\ref{angular02}), suggesting compression in the angular direction but up to this finite value. While in the case of Schwarzschild black hole the angular tidal force diverges to negative infinity as $r\rightarrow{}0$ which means infinite compression in the angular direction for a radially free-falling object.

In Fig. \ref{tidal_comp}, we compare the radial tidal force components of null naked singularity and Schwarzschild black hole spacetimes. In this plot, we have considered the ratio in the cases of different $ADM$ mass values i.e. $M=(1, 5, 10, 20, 30, 100)$. It is particularly evident from this plot that the ratio approaches zero near the singularity, indicating the fact that radial tidal force in null naked singularity settles to a finite value beyond its maxima and it diverges in the case of Schwarzschild black hole, resulting the ratio approaching zero near the singularity. The profile of all the curves in this figure approaches unity for an asymptotical distance from the singularity and only near the maximal distance for each particular singularity i.e. at $M=r$, the ratio begins to rapidly converge to zero.

\section{Geodesic deviation}
\label{V}
Substituting the explicit form of $f(r)$ in null naked singularity spacetime into Eqs.~(\ref{radial}) and~(\ref{angular}), we obtain the tidal force equations for our spacetime to be:
\begin{equation}
 \left(E^{2}-\left(1+\frac{M}{r}\right)^{-2}\right){\eta}^{\widehat{1}}{''}=\frac{Mr}{(M+r)^{3}}{\eta}^{\widehat{1}}{'}-\frac{M(M-2r)}{(M+r)^{4}}\eta^{\widehat{1}},  
 \label{5.1}
\end{equation}
\begin{equation}
 \left(E^{2}-\left(1+\frac{M}{r}\right)^{-2}\right){\eta}^{\widehat{i}}{''}=\frac{Mr}{(M+r)^{3}}{\eta}^{\widehat{i}}{'}-\frac{M}{(M+r)^{3}}\eta^{\widehat{i}}.
 \label{5.2}
\end{equation}

From the above two equations, it can be inferred that the tidal forces in this spacetime depend on the mass and the radial distance from the singularity. We note that the expressions of the tidal forces, given by Eqs.~(\ref{radial}) and~(\ref{angular}), are identical to the Newtonian tidal forces with the force $-f'/2$ in the radial direction. Using the above equations (\ref{5.1}) and (\ref{5.2}), we compute their numerical solutions for $\eta^{\hat 1}(r)$ and $\eta^{\hat i}(r)$ as depicted in Fig. \ref{geodesic}. Both the plots Fig. \ref{geodesic}.(a) and \ref{geodesic}.(b) represent how the initial radial and angular separations, respectively, of two nearby geodesics are changing while falling towards the central singularity in both the cases of a null naked singularity and Schwarzschild black hole. It is appropriate to comment that
$\eta^{\hat 1}(b)$ and $\eta^{\hat i}(b)$ are initial separation
distances between two nearby geodesics at $r=b$ to the radial and angular directions, respectively, and in the case of both the plots we have considered the impact parameter $b = 100 M$. Additionally, $\frac{d\eta^{\hat
1}(b)}{d\tau}$ and $\frac{d\eta^{\hat i}(b)}{d\tau}$ are initial velocities at $r=b$ to the radial and angular directions, respectively. We have considered the case when $\eta^{\hat\alpha}(b)\neq 0$,
$\frac{d\eta^{\hat \alpha}(b)}{d\tau}=0$, where $(\alpha=1,~2,~3)$.
 
It is evident from Fig. \ref{geodesic}.(a) that the radial component of the geodesic separation vector in both the cases i.e. for null naked singularity and Schwarzschild black hole follow a similar trajectory. This implies that the geodesic separation vector $\eta^{\widehat{1}}$ gets stretched radially and approaches infinity as the separation vector comes closer to the singularity. In the case of angular component of the geodesic separation vector $\eta^{\widehat{i}}$, Fig. \ref{geodesic}.(b) illustrates the act of compression as the separation vector tends to zero when near the singularity and the profile of the plot is the same in both null naked and Schwarzschild singularities. 

\section{Conclusions}
\label{VI}
 The comparison of null naked singularity spacetime and the Schwarzschild black hole spacetime was intriguing not only from the very inception of the idea for the same but instead, it concluded with some very interesting results. The important remarks and conclusions of this work are as follows:
\begin{itemize}
    \item The orbit's precession angle in Schwarzschild spacetime is greater than that of the naked singularity spacetime orbit for the same parameters value. However, as the angular momentum of the particle 'h' increases, the precession angle of the orbit becomes almost equivalent at larger distances from the singularity. As we point out before, the null naked singularity asymptotically resembles the Schwarzschild spacetime, but the nature of the orbital precession remains unchanged. As a result, the null naked singularity can be thought of as a Schwarzschild black hole mimicker.
    
    \item The most counter-intuitive result that we were able to obtain is the fact that the radial component of the tidal force in null naked singularity spacetime also shows a steep rise but, at a later stage when compared to the case of Schwarzschild black hole as shown in Fig. \ref{tidal_plt}(a). And this steep rise in the radial tidal force does not culminate into an infinite force when approaching the singularity like in the case of the Schwarzschild black hole but instead it flips its signature by turning from a stretching force to a compressing force at a finite distance from the singularity. The magnitude of which would depend on the inverse square of the ADM mass $M$ of the spacetime i.e. $-{1}/{M^2}$ when $r \rightarrow{}0$ from Eq. (\ref{radial02}). This aspect alone can be sufficient to distinguish between a null naked singularity spacetime and a Schwarzschild spacetime. 
    \item To add further we have also compared the results for the angular component of the tidal force for the two spacetimes and a similar steep rise at a later stage in the angular tidal force is seen in the case of null naked singularity, but no change in the signature of the tidal force component. Only the fact that the magnitude of the angular part of the tidal force in null naked singularity approaches $-{1}/{M^2}$ as $r \rightarrow{}0$ from Eq. (\ref{angular02}), which suggests a finite force of compression which is dependent only on the inverse square of the ADM mass $M$ of the spacetime. This is another distinguishing aspect between the two spacetimes, as in the Schwarzschild spacetime the angular component of the tidal force is solely dependent on the inverse cube of the radial distance from the singularity \cite{Hong:2020bdb}.
    
    \item We also computed the numerical solution of the geodesic deviation equations about a radially free-falling geodesic. In particular, we examined the behavior of the geodesic separation vector for such a geodesic
under the influence of tidal forces created by the null naked singularity spacetime and compared the results to Schwarzschild black hole spacetime. We conclude that both spacetimes show similar behavior of the geodesic separation vector.

\end{itemize}

\section{Acknowledgments}
The authors would like to acknowledge the support and encouragement provided by all the members and research fellows of ICC (International Centre for Cosmology). Also would like to express our gratitude towards Pankaj S. Joshi for all his valuable suggestions and astute insights.
\section*{Data Availability Statement}
This manuscript has no associated data
or the data will not be deposited. [Authors’ comment: The results were
obtained via analytical and numerical simulations. So there are no associated data.]

\end{document}